\documentclass[journal,12pt,draftclsnofoot,onecolumn]{IEEEtran}

\ifCLASSINFOpdf
\else

\fi

\usepackage{doi}
\usepackage{tikz}
\usepackage{graphicx}
\usepackage{mathtools}
\usepackage{romannum}
\usepackage{amsmath,amsfonts,amsthm,amssymb,mathrsfs}
\usepackage{type1cm}
\usepackage{lineno}
\usepackage{eso-pic}
\usepackage{color}
\usepackage{subfigure}
\usepackage{color}
\usepackage{fancyhdr}
\usepackage{lastpage}
\usepackage{adjustbox}
\usepackage{xfrac}
\DeclareGraphicsExtensions{.pdf,.jpeg,.png}
\graphicspath{{./img/}}

\begin{document}

\title{Non-Iterative Subspace-Based DOA Estimation in the Presence of Nonuniform Noise}

\author{Majdoddin~Esfandiari, 
	    Sergiy~A.~Vorobyov,~\IEEEmembership{Fellow,~IEEE},
        Simin~Alibani, 
        and~Mahmood~Karimi 
\thanks{M. Esfandiari and S. A. Vorobyov are with Dept. Signal Processing and Acoustics, Aalto University, PO Box 15400, 00076 Aalto, Finland. S. Alibani and M. Karimi are with School of Electrical and Computer Engineering, Shiraz University, Shiraz, Iran.
Emails: {majdoddin.esfandiari@aalto.fi; sergiy.vorobyov@aalto.fi; alibani.simin@gmail.com; karimi@shirazu.ac.ir}.
Corresponding author is S. A. Vorobyov.
}} 

\markboth{IEEE Signal Processing Letters}%
{Shell \MakeLowercase{\textit{et al.}}: Bare Demo of IEEEtran.cls for IEEE Journals}


\maketitle

\begin{abstract}
The uniform white noise assumption is one of the basic assumptions in most of the existing directional-of-arrival (DOA) estimation methods. 
In many applications, however, the non-uniform white noise model is more adequate. Then the noise variances at different sensors have to be also estimated as nuisance parameters while estimating DOAs. In this letter, different from the existing iterative methods that address the problem of non-uniform noise, a non-iterative two-phase subspace-based DOA estimation method is proposed. The first phase of the method is based on estimating the noise subspace via eigendecomposition (ED) of some properly designed matrix and it avoids estimating the noise covariance matrix. In the second phase, the results achieved in the first phase are used to estimate the noise covariance matrix, followed by estimating the noise subspace via generalized ED. Since the proposed method estimates DOAs in a non-iterative manner, it is computationally more efficient and has no convergence issues as compared to the existing methods. Simulation results demonstrate better performance of the proposed method as compared to other existing state-of-the-art methods.
\end{abstract}


\IEEEpeerreviewmaketitle

\section{Introduction}
\IEEEPARstart{D}{irection} of Arrival (DOA) and spectral estimation are the fundamental problems in array processing and spectral analysis with many applications in radar, sonar, navigation and communication systems, as well as acoustic tracking to mention just a few \cite{r14a}--\cite{r15}. There exist several DOA estimation approaches. Among the most notable are the maximum likelihood (ML), beamforming-based, parametric subspace-based, and sparse representation-based approaches \cite{r1}--\cite{r8}. Parametric subspace-based DOA estimation methods, such as multiple signal classification (MUSIC) \cite{r1}, \cite{r6} and estimation of signal
parameters via rotational invariance technique (ESPRIT) \cite{r7} are well known to provide high accuracy and high resolution for DOA estimation with low computational complexity in comparison to the methods such as ML \cite{r1}. In addition to the traditional far-field narrow band signal assumption, the assumption of uncorrelated sources is also critical for the former methods. A fundamental assumption that applies to all the aforementioned methods is however the presence of spatially uniform white noise. Under this assumption, the analytic concentration of the ML function with respect to the noise variance single parameter becomes possible, while parametric subspace-based methods are just built on this assumption since it enables separation of signal and noise subspaces \cite{Mahdi}, \cite{Matt}. 

In diverse practical scenarios, the spatially uniform white noise assumption may be violated. Indeed, the sensor noise may be  non-uniform \cite{r2}, \cite{r3}--\cite{r12a}, spatially correlated \cite{r9}, \cite{r10}, or block-correlated \cite{r2a}. Spatially white non-uniform noise arises when sensor noise powers are non-identical across the array, and leads to diagonal noise covariance matrix with not-identical entries. 
To overcome the problem of performance degradation in the presence of non-uniform noise, a variety of DOA estimation algorithms and techniques have been proposed. In \cite{r2} and \cite{r3}, two deterministic and stochastic ML estimators are respectively proposed based on iterative procedures. These two estimators suffer from high computational complexity. Thus, two iterative subspace estimation algorithms with lower complexity, called iterative maximum likelihood subspace estimation (IMLSE) and iterative least squares subspace estimation (ILSSE), respectively, based on ML and least squares (LS) have been proposed in \cite{r5} for estimating signal subspace and noise covariance matrix. These algorithms then use spectral MUSIC method for performing the DOA estimation. 

In this letter, we propose a new subspace-based method for DOA estimation in spatially non-uniform noise, which is non-iterative, thus leading to lower computational complexity and avoiding any convergence issues. The method has two phases. In the first phase, the noise subspace is initially estimated via eigendecomposition (ED) of some properly designed matrix without knowing the noise covariance matrix. In the second phase, the noise covariance matrix is then estimated by exploiting the results of the first phase and then the generalized ED is applied to the output array covariance matrix and noise covariance matrix. Simulation results demonstrate the efficiency and superiority of the proposed method in terms of both performance and complexity over the existing methods.

\section{SIGNAL MODEL}
Consider an array of $M$ sensors receiving $L$ ($L<M$ is known) independent narrowband signals impinging from the sources in far-field. The signal observed at time instance $t$ by the array is given as
\begin{flalign}
\mathbf{x}(t)=\mathbf{A}(\boldsymbol{\theta})\mathbf{s}(t)+\mathbf{n}(t)
 \label{eq1} \
\end{flalign} 
where $\mathbf{A} (\boldsymbol{\theta}) \triangleq [\mathbf{a}(\theta_1), \cdots, \mathbf{a}(\theta_L)]$ is the $M\times L$ matrix whose columns are the signal steering vectors $\mathbf{a}(\theta_i),$ $i=1,\cdots, L$, $\boldsymbol{\theta} \triangleq [\theta_1, \cdots, \theta_L]^T$ is the vector of unknown source DOA's, $\mathbf{s}(t)$ is the $L\times 1$ vector of source signals, $\mathbf{n}(t)$ is the $M\times 1$ vector of zero-mean spatially and temporally white Gaussian noise, $N$ is the number of snapshots, $t$ is the discrete time index, and $[ \cdot ]^T$ denotes the transpose. 

Using \eqref{eq1}, the array output covariance matrix can be expressed as
\begin{flalign}
\mathbf{R} \triangleq E\{\mathbf{x}(t) \mathbf{x}^H(t)\} = \mathbf{A} (\boldsymbol{\theta}) \mathbf{P} \mathbf{A}^H (\boldsymbol{\theta}) + \mathbf{Q}
\label{eq2} \
\end{flalign} 
where $\mathbf{P} \triangleq E\{\mathbf{s}(t)\mathbf{s}^H(t)\}$ and $\mathbf{Q} \triangleq E\{\mathbf{n}(t)\mathbf{n}^H(t)\}$ are respectively the $L\times L$ signal and $M\times M$ noise covariance matrices, and $E\{\cdot\}$ and $(\cdot)^H$  denote the expectation and Hermitian transpose operators, respectively. For uncorrelated sources, $\mathbf{P}$ is a diagonal matrix. Under the non-uniform uncorrelated noise assumption, $\mathbf{Q}$ is also a diagonal matrix of the form
\begin{flalign}
\mathbf{Q}=diag \left\{ \sigma_1^2,\cdots, \sigma_M^2 \right\} 
\label{eq4} \
\end{flalign} 
where $\sigma_m^2$, $m=1, \cdots, M$ are the sensor noise variances and $diag\{ \cdot\}$ denotes a diagonal matrix.

\section{NEW PROPOSED METHOD}
In parametric subspace-based DOA estimation, the noise subspace needs to be first estimated. Under the non-uniform noise, also the noise covariance matrix possibly needs to be estimated. Then the orthogonality of the noise subspace basis vectors and the source steering vectors can be used to estimate the source DOA's.  


To estimate the noise subspace, recall that in \eqref{eq2}, $\mathbf{A}^H(\boldsymbol{\theta})$ is an $L\times M$ full row-rank matrix, and there are $M-L$ orthonormal vectors $\mathbf{u}_l$, $l=1,\cdots, M-L$ satisfying the following homogenous equation
\begin{flalign}
\mathbf{A}^H(\boldsymbol{\theta})\mathbf{u}_l=\mathbf{0}
\label{eq5} \
\end{flalign}     
where $\mathbf{0}$ is the vector of zeros. Multiplying both sides of \eqref{eq2} on the right by $\mathbf{u}_l$ and using \eqref{eq5}, we obtain 
\begin{flalign}
\mathbf{R}\mathbf{u}_l = \mathbf{A} (\boldsymbol{\theta}) \mathbf{P} \mathbf{A}^H (\boldsymbol{\theta}) \mathbf{u}_l + \mathbf{Q} \mathbf{u}_l = \mathbf{Q} \mathbf{u}_l, \ \ l=1,\cdots, M-L.
\label{eq6}
\end{flalign}  
According to \eqref{eq6}, $M-L$ vectors $\mathbf{u}_l$, $l=1,\cdots, M-L$ which span the range space of the noise subspace are the solutions of the generalized ED problem for the matrices $\mathbf{R}$ and $\mathbf{Q}$, while all $M-L$ eigenvalues being equal to one. However, since the matrix $\mathbf{Q}$ is unknown, $\mathbf{u}_l$, $l=1,\cdots, M-L$ cannot be computed as simple as in the uniform noise case. 

We observe, however, that the array output covariance matrix $\mathbf{R}$ can be written as the following sum of two matrices
\begin{flalign}
\mathbf{R} = \mathbf{R}_1+\mathbf{R}_2
\label{eq7} \
\end{flalign}  
where
\begin{flalign}
[\mathbf{R}_1]_{i,j} = \begin{cases}
[\mathbf{R}]_{i,j}, & \quad i\neq j \\
0, & \quad i=j
\end{cases}
\label{eq8} \
\end{flalign}
and
\begin{flalign}
\mathbf{R}_2& = diag \left\{ [\mathbf{R}]_{1,1}, \cdots, [\mathbf{R}]_{M,M} \right\} \nonumber \\
&=diag \left\{ \displaystyle\sum_{k=1}^{L} s_k + \sigma_1^2, \cdots, \displaystyle\sum_{k=1}^{L} s_k +\sigma_M^2 \right\} 
\label{eq9} \
\end{flalign}
with $s_k$ being the received power of the $k$th source. 

Substituting \eqref{eq7}, \eqref{eq8}, and \eqref{eq9} into \eqref{eq6}, we obtain
\begin{flalign}
\mathbf{R}_1\mathbf{u}_l = (\mathbf{Q} - \mathbf{R}_2) \mathbf{u}_l = - \left( \displaystyle\sum_{k=1}^{L} s_k \right) \mathbf{u}_l .
\label{eq10} \
\end{flalign}
It can be seen from \eqref{eq10} that $\mathbf{u}_l$, $l=1,\cdots, M-L$ can be obtained by applying ED to the matrix $\mathbf{R}_1$ only. As a matter of fact, since adding the scaled identity matrix of the form $b \cdot \mathbf{I}$ to $\mathbf{R}_1$ does not alter eigenvectors and only shifts eigenvalues of $\mathbf{R}_1$, the noise subspace basis vectors $\mathbf{u}_l$, $l=1,\cdots, M-L$ can be computed by applying ED to every matrix whose diagonal elements are identical and off-diagonal elements are equal to the off-diagonal elements of $\mathbf{R}$. Thus, after applying ED to $\mathbf{R}_1$, the noise subspace basis vectors ${\mathbf{u}}_l$, $l=1,\cdots, M-L$, or in matrix form $\mathbf{U} \triangleq [\mathbf{u}_1,\cdots,\mathbf{u}_{M-L}]$, can be obtained even without the need to estimate $\mathbf{Q}$. This novel result is stated and proved in the following lemma. 

{\bf Lemma 1:} The noise subspace basis vectors ${\mathbf{u}}_l$, $l=1,\cdots, M-L$ are the $M-L$ eigenvectors of ${\mathbf{R}}_1$ whose corresponding eigenvalues are the smallest.

{\it Proof:} Assume that there exists an $M \times 1$ vector $\mathbf{d}$ that satisfies the following conditions
\begin{flalign}
& \mathbf{A}(\boldsymbol{\theta})\mathbf{P}\mathbf{A}^H(\boldsymbol{\theta})\mathbf{d}\neq\mathbf{0} \label{eq21} \\ 
& \mathbf{R}_1\mathbf{d}=\lambda\mathbf{d} .
\label{eq22} \
\end{flalign}
Adding $\mathbf{R}_2 \mathbf{d}$ to both sides of \eqref{eq22} and using \eqref{eq7}, we obtain
\begin{flalign}
\mathbf{R}_1\mathbf{d}+\mathbf{R}_2\mathbf{d}=\mathbf{R}\mathbf{d}=\lambda\mathbf{d}+\mathbf{R}_2\mathbf{d}
\label{eq23}. \
\end{flalign} 

Inserting \eqref{eq2} into \eqref{eq23}, and rearranging the terms, we have
\begin{flalign}
\mathbf{A}(\boldsymbol{\theta})\mathbf{P}\mathbf{A}^H(\boldsymbol{\theta})\mathbf{d}=\lambda\mathbf{d}+(\mathbf{R}_2-\mathbf{Q})\mathbf{d} .
\label{eq24} \
\end{flalign}
Moreover, substituting \eqref{eq4} and \eqref{eq9} into \eqref{eq24}, we obtain
\begin{flalign}
\mathbf{A}(\boldsymbol{\theta})\mathbf{P}\mathbf{A}^H(\boldsymbol{\theta})\mathbf{d} = \left( \lambda + \displaystyle \sum_{k=1}^{L}  s_k \right) \mathbf{d} .
\label{eq25} \
\end{flalign}

It can be seen from \eqref{eq25} that $\mathbf{d}$ is an eigenvector of the matrix $\mathbf{A} (\boldsymbol{\theta}) \mathbf{P} \mathbf{A}^H (\boldsymbol{\theta})$ while its corresponding eigenvalue is equal to $\lambda + \sum_{k=1}^{L} s_k$. Since the matrix $\mathbf{A} (\boldsymbol{\theta}) \mathbf{P} \mathbf{A}^H (\boldsymbol{\theta})$ has $L$ positive eigenvalues and the condition \eqref{eq21} has to be satisfied, it can be concluded that
\begin{flalign}
\lambda + \displaystyle\sum_{k=1}^{L} s_k > 0 \quad \Rightarrow \quad \lambda>-\sum_{k=1}^{L} s_k .
\label{eq26} \
\end{flalign}
In other words, \eqref{eq26} indicates that $-\sum_{k=1}^{L} s_k$ is the lower bound on the smallest eigenvalue of $\mathbf{R}_1$. Thus, the noise subspace basis ${\mathbf{u}}_l$, $l=1,\cdots, M-L$ is composed of $M-L$ eigenvectors of $\mathbf{R}_1$ with the smallest eigenvalues. $\square$

Knowing $\mathbf{U}$, the spectral-MUSIC method, for example, can be used for the source DOA's estimation by finding the locations of $L$ peaks in the pseudo-spectrum
\begin{flalign}
\mathbf{S}(\theta)=\frac{1}{\mathbf{a}^H(\theta)\mathbf{U}\mathbf{U}^H\mathbf{a}(\theta)} .
\label{eq11} \
\end{flalign}
However, the estimate of $\mathbf{U}$ can be further improved using the results of the initial $\mathbf{U}$ estimation and	exploiting \eqref{eq6}. Indeed, if there exists an estimate of $\mathbf{Q}$, \eqref{eq6} can be solved by applying generalized ED to the matrices $\mathbf{R}$ and $\mathbf{Q}$. Then more accurate noise subspace basis vectors $\mathbf{u}_l$, $l=1,\cdots, M-L$ can be found as stated in the following lemma.

{\bf Lemma 2:} The noise subspace basis vectors ${\mathbf{u}}_l$, $l=1,\cdots, M-L$ are the $M-L$ eigenvectors, obtained by applying generalized eigendecomposion to the matrices ${\mathbf{R}}$ and $\mathbf{Q}$ whose corresponding eigenvalues are the smallest.

{\it Proof:} Similar to the proof of Lemma~1, assume that there exists an $M \times 1$ vector $\mathbf{d}$ that satisfies \eqref{eq21} and $\mathbf{R}\mathbf{d}=\lambda\mathbf{Q}\mathbf{d}$. 

Inserting \eqref{eq2} into $\mathbf{R}\mathbf{d}=\lambda\mathbf{Q}\mathbf{d}$, and rearranging the terms, we have
\begin{flalign}
\mathbf{A}(\boldsymbol{\theta})\mathbf{P}\mathbf{A}^H(\boldsymbol{\theta})\mathbf{d}=(\lambda-1)\mathbf{Q}\mathbf{d} .
\label{eq29} \
\end{flalign}
Multiplying both sides of \eqref{eq29} on the left by $\mathbf{d}^H$, we obtain
\begin{flalign}
\mathbf{d}^H\mathbf{A}(\boldsymbol{\theta})\mathbf{P}\mathbf{A}^H(\boldsymbol{\theta})\mathbf{d}=\bar{\mathbf{d}}^H\mathbf{P}\bar{\mathbf{d}}=(\lambda-1)\mathbf{d}^H\mathbf{Q}\mathbf{d} 
\label{eq30} \
\end{flalign}
where $\bar{\mathbf{d}}=\mathbf{A}^H(\boldsymbol{\theta})\mathbf{d}$. Since both $\mathbf{P}$ and $\mathbf{Q}$ are positive definite matrices and the condition \eqref{eq21} has to be satisfied, it can be concluded that
\begin{flalign}
\lambda - 1 > 0 \quad \Rightarrow \quad \lambda>1 .
\label{eq31} \
\end{flalign}
In other words, \eqref{eq31} indicates that $1$ is the lower bound on the smallest eigenvalue of generalized ED of $\mathbf{R}$ and $\mathbf{Q}$. Thus, the noise subspace basis ${\mathbf{u}}_l$, $l=1,\cdots, M-L$ is composed of $M-L$ eigenvectors with the smallest eigenvalues. $\square$

To apply Lemma~2, we first need to find an estimate of $\mathbf{Q}$. Towards this end, first we write $\mathbf{Q}$ as the following sum of two diagonal matrices
\begin{flalign}
\mathbf{Q} = \sigma^2 \mathbf{I} + \mathbf{Q}_{\rm nun}
\label{eq12} \
\end{flalign}
where $\sigma^2$ represents the common part of sensor noise powers, which is computed later, and $\mathbf{Q}_{\rm nun}$ is a diagonal matrix whose diagonal elements, except for one of them, are nonzero. The place of this zero element is the place of the smallest diagonal element of $\mathbf{R}$. As a result, the rank of $\mathbf{Q}_{\rm nun}$ is $M-1$, and  
\begin{flalign}
\mathbf{e}^T_k \mathbf{Q}_{\rm nun} = \mathbf{0} 
\label{eq13} 
\end{flalign}
where $\mathbf{e}_k$ is the $M \times 1$ unit vector such that 
\begin{flalign}
[\mathbf{e}_k]_i = \begin{cases}
0, & \quad i\neq k \\
1, & \quad i=k
\end{cases}
\label{eq40}
\end{flalign}
and $k$ is the index of the smallest diagonal element of $\mathbf{R}$. 

Multiplying both sides of \eqref{eq6} by $\mathbf{e}_k^T$ on the left and using \eqref{eq12} and \eqref{eq13}, we obtain
\begin{flalign}
\mathbf{e}_k^T \mathbf{R} \mathbf{u}_l = \mathbf{e}_k^T (\sigma^2 \mathbf{I} + \mathbf{Q}_{\rm nun}) = \sigma^2 \mathbf{e}_k^T \mathbf{u}_l .
\label{eq14} \
\end{flalign}
Equation \eqref{eq14} can be written for all vectors $\mathbf{u}_l$, $l=1,\cdots, M-L$ in the following matrix-vector form
\begin{flalign}
\mathbf{e}_k^T \mathbf{R} \mathbf{U} = \sigma^2 \mathbf{e}_k^T \mathbf{U}
\label{eq15} \
\end{flalign}
where $\mathbf{U}$ is composed of $\mathbf{u}_l$, $l=1,\cdots, M-L$ obtained by Algorithm~1. Consequently, $\sigma^2$ can be computed as
\begin{flalign}
\sigma^2 = \frac{\mathbf{e}_k^T \mathbf{R} \mathbf{U} \mathbf{U}^H \mathbf{e}_k}{\mathbf{e}_k^T \mathbf{U} \mathbf{U}^H \mathbf{e}_k} .
\label{eq16} \
\end{flalign}

The only issue remaining is the construction of the matrix $\mathbf{Q}_{\rm nun}$. Let us set the nonzero diagonal elements of $\mathbf{Q}_{\rm nun}$ as the differences of the corresponding elements in $\mathbf{R}$ with the smallest diagonal element of $\mathbf{R}$, that is,
\begin{flalign}
\mathbf{Q}_{\rm nun} = diag \left\{ [\mathbf{R}]_{1,1} - c, \cdots, [\mathbf{R}]_{M,M} - c \right\}
\label{eq17} \
\end{flalign} 
where $c$ is the smallest diagonal element of $\mathbf{R}$.

Finally, the matrix $\mathbf{Q}$ can be formed by utilizing \eqref{eq12}, \eqref{eq16}, and \eqref{eq17}. With the matrices $\mathbf{R}$ and $\mathbf{Q}$, the noise subspace basis ${\mathbf{u}}_l$, $l=1,\cdots, M-L$ can be re-estimated as stated in \eqref{eq6} and Lemma~2. Then the new $\mathbf{U}$ can be formed and the source DOA's can be estimated by finding, for example, the locations of $L$ peaks in \eqref{eq11}. The corresponding algorithm for DOA estimation in non-uniform noise is summarized in Algorithm~1, where the sample data covariance matrix $\hat{\mathbf{R}}$ is used as an estimate of the array output covariance matrix $\mathbf{R}$. Steps 1 and 2 represent the first phase of the algorithm that can be followed by step 7 directly. Steps 3--6 represent the second {\it correction} phase of the algorithm.

\begin{table}[!h]
	\label{tb4}
		\begin{tabular}{l}
			\hline
			\textbf{Algorithm 1:} The proposed method.
		\\ \hline
		1: Compute the sample covariance matrix $\hat{\mathbf{R}} = \frac{1}{N}\sum_{t=1}^{N} \mathbf{X}(t)\mathbf{X}^H(t)$. \\
		2: Form $\hat{\mathbf{R}}_1$ from $\hat{\mathbf{R}}$ as in \eqref{eq8}, carry out the ED of $\hat{\mathbf{R}}_1$ to obtain the \\  noise subspace basis $\hat{\mathbf{u}}_l$, $l=1,\cdots, M-L$, and construct the matrix $\hat{\mathbf{U}}$. \\
		3: Construct $\mathbf{e}_k$ and $\hat{\mathbf{Q}}_{\rm nun}$ according to \eqref{eq13}, \eqref{eq40}, and \eqref{eq17}. \\
		4: Using the data sample covariance matrix and $\hat{\mathbf{U}}$ obtained in step 2, \\ 
		estimate $\hat{\sigma}^2$ according to \eqref{eq16}. \\
		5: Using, $\hat{\sigma}^2$ and $\hat{\mathbf{Q}}_{\rm nun}$, estimate $\hat{\mathbf{Q}}$ according to \eqref{eq12}.\\
		6: Apply generalized ED to $\hat{\mathbf{R}}$ and $\hat{\mathbf{Q}}$, and obtain the new estimate of the \\ noise subspace basis $\hat{\mathbf{u}}_l$, $l=1,\cdots, M-L$, i.e., the new estimate of $\hat{\mathbf{U}}$. \\
		7: Use spectral-MUSIC, i.e., find the locations of $L$ peaks in \eqref{eq11}, where \\ $\mathbf{U}$ is substituted its estimate $\hat{\mathbf{U}}$. \\
		\hline
		\end{tabular}
 \end{table}

\textit{Complexity analysis:} For the proposed method, the ED of $\mathbf{R}_1$ or the generalized ED of $\mathbf{R}$ and $\mathbf{Q}$ are involved. The corresponding complexity is \textit{O}($M^3$) \cite{r20}. It is equivalent to the complexity in each iteration of IMLSE or ILSSE \cite{r5}. The difference arises from the fact that IMLSE and ILSSE are iterative methods (the number of iterations can be comparable to $M$ in many scenarios to converge to their best result).

\section{SIMULATION RESULTS}
A ULA with $M=8$ omnidirectional sensors, which are separated by half wavelength, is considered. Two far-field uncorrelated narrow-band signals (with equal powers) impinge on the array simultaneously from $\theta_1=-3^{\circ}$ and $\theta_2=6^{\circ}$, respectively.  

The worst noise power ratio (WNPR), and the signal-to-noise ratio (SNR) are defined as
${\rm WNPR} = \sfrac{\sigma^2_{\rm max}}{\sigma^2_{\rm min}}$ and 
${\rm SNR} = \sfrac{\sigma^2_{\rm s}}{M} \sum_{m=1}^{M} (\sigma^2_m)^{-1}$, respectively, where $\sigma^2_{\rm max}$ and $\sigma^2_{\rm min}$ are the maximum and minimum sensors noise powers, respectively, and $\sigma^2_{\rm s}$ is the signal power. The number of snapshots ($N$) and the number of Monte Carlo trials ($K$) are set to 500 and 5000, respectively. The root mean squared error (RMSE) of DOA's estimation is defined as
\begin{flalign}
{\rm RMSE} = \sqrt{\frac{1}{KL}\displaystyle\sum_{k=1}^{K} \displaystyle\sum_{l=1}^{L} (\hat{\theta}_{k,l}-\theta_l)^2}
\nonumber
\end{flalign}
where $\hat{\theta}_{k,l}$ denotes the $l$th DOA estimate in the $k$th trial. 

To validate the performance of the proposed methods, two examples are considered, and the results are compared to the performance of the standard spectral-MUSIC and well as IMLSE and ILSSE methods both after the first iteration only and also after convergence is achieved.\footnote{The performance of ILSSE is plotted only for the SNRs when ILSSE converges for all trials.} The number of sources is assumed to be known for all methods tested.

\begin{figure}[!]
	\begin{center}
		{\includegraphics[width=5.5in,height=5.5in]{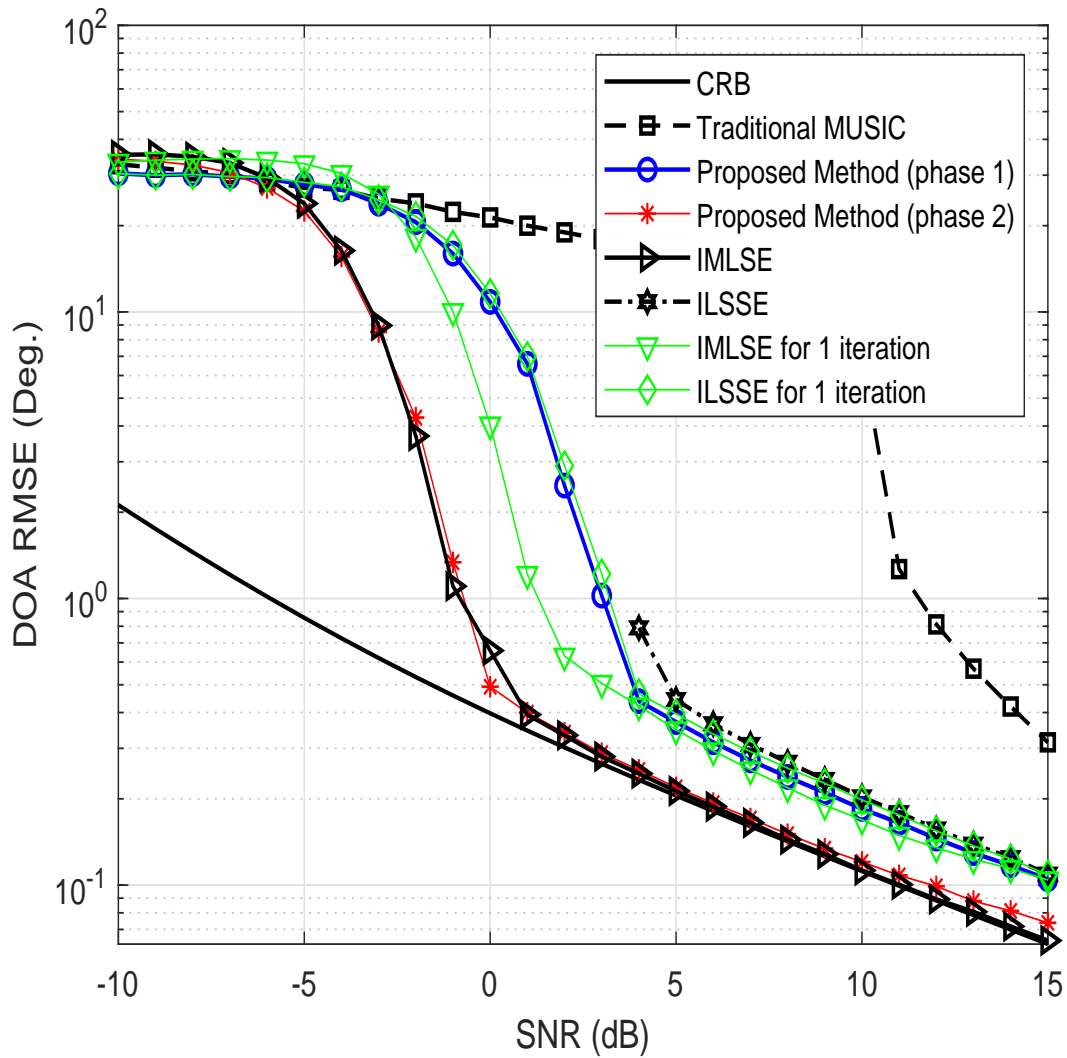}}
		\caption{The RMSEs of DOA estimation versus SNR in Example~1.}
		\label{fig1}
	\end{center}
\end{figure}

{\it Example 1:} The non-uniform noise covariance matrix is fixed in all simulation runs and is given as $\mathbf{Q}=diag\{1, 1, 1, 1, 1, 20, 30, 50\}$, resulting in WNPR=50. Fig.~\ref{fig1} shows the RMSEs for the methods tested versus SNR. The Cramer-Rao bound (CRB) \cite{r2} is also shown. It can be seen from the figure that the proposed method after both phases demonstrates the best threshold behavior, despite it is non-iterative and has lower computational complexity. Although the IMLSE method after the first iteration only possesses as low computational complexity as the proposed method, its DOA estimation accuracy is very poor. Thus, multiple iterations have to run for it to achieve its best result.

\begin{figure}[]
	\begin{center}
		{\includegraphics[width=5.5in,height=5.5in]{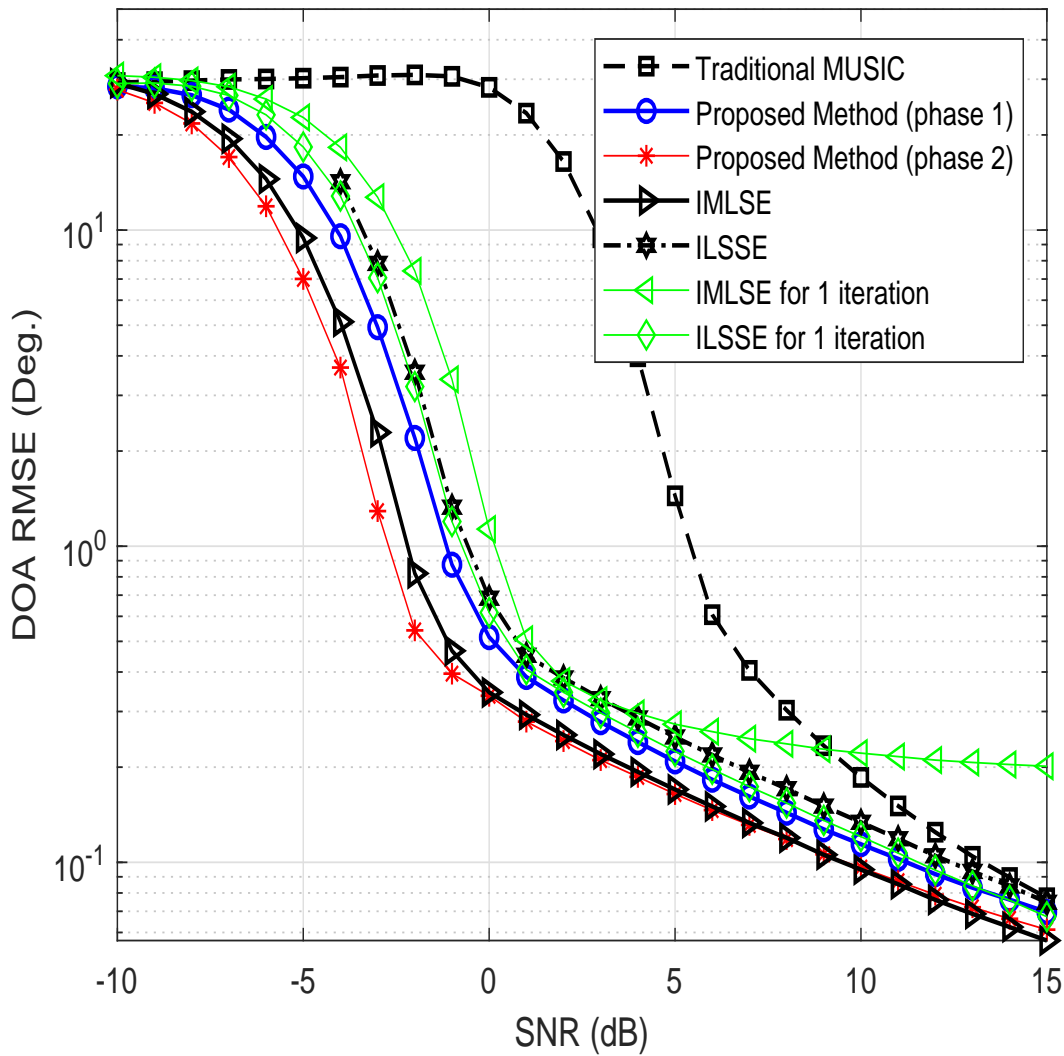}}
		\caption{The RMSEs of DOA estimation versus SNR in Example~2.}
		\label{fig3}
	\end{center}
\end{figure}  

{\it Example 2:} To provide more comprehensive insights into the performance of the methods tested, the noise covariance matrix is chosen in this example multiple times randomly with maximum WNPR of 30. Then the RMSE results are also averaged over 50 different realizations of the noise covariance matrix $\mathbf{Q}$ for which of each 5000 Monte Carlo trials are averaged. The average RMSEs for the methods tested are shown in Fig.~\ref{fig3}.\footnote{Since the curves in Fig.~2 are resulted also from averaging over different realization of the nuisance parameters, i.e., $\mathbf{Q}$'s with different WNPRs, the CRB is not applicable in the setup and is not shown. Indeed, fixed parameters of interest as well as nuisance parameters have to be assumed for CRB.} It can be seen from the figure that the second phase of the proposed method shows the best performance and improves the threshold behavior by about 2~dB as compared to the next best performing method that is the IMLSE method. 

Furthermore, Table~I shows the average run time of the methods tested for SNRs=-5, 0, 5, 10 , and 15~dB for the setup of Example~2. The simulation is performed on a PC running an Intel(R) Xeon(R) 3.40GHz CPU. It can be observed that the proposed method is superior in terms of the required time which is reduced by orders of magnitude compared to the existing methods. The average number of iterations for IMLSE and ILSSE are about 27 and 52, respectively, for SNR=5~dB, for example. Inspecting Fig.~2 and Table~I together, it can be seen that even the first phase of the proposed method leads to superior performance with lower complexity compared to the competitive methods after the first iteration.

\begin{table}[!h]
	\label{tb2}
	\caption{Comparison of average run time of one trial (in $ms$).}
	\begin{adjustbox}{width=1.\textwidth, height=0.09\textheight}
		\begin{tabular}{lllllll}
			\hline
			\textbf{Method}& IMLSE& \ \ \ ILSSE& IMLSE after& ILSSE after& Proposed method& Proposed method  \\
			&  & &first iteration& first iteration& \ \ \ (phase 1)& \ \ \ (phase 2)  
			\\ \hline
			SNR=-5~dB& 13.108& Not converged& \ \ \ 0.573& \ \ \ 0.478& \ \ \ \ \ 0.371& \ \ \ \ \ 0.891 \\
			\hline
			SNR=0~dB& 14.395& \ \ \ 21.739& \ \ \ 0.552& \ \ \ 0.470& \ \ \ \ \ 0.877& \ \ \ \ \ 0.877 \\
			\hline
			SNR=5~dB& 15.133& \ \ \ 23.870& \ \ \ 0.546& \ \ \ 0.462& \ \ \ \ \ 0.359& \ \ \ \ \ 0.861 \\
			\hline
			SNR=10~dB& 15.470& \ \ \ 26.667& \ \ \ 0.545& \ \ \ 0.459& \ \ \ \ \ 0.357& \ \ \ \ \ 0.860 \\
			\hline
			SNR=15~dB& 15.604& \ \ \ 26.385& \ \ \ 0.547& \ \ \ 0.459& \ \ \ \ \ 0.357& \ \ \ \ \ 0.861 \\
			\hline
		\end{tabular}
	\end{adjustbox} 
\end{table}

\section{Conclusion}
A novel computationally efficient non-iterative two-phase subspace-based parametric method for DOA estimation in the presence of unknown spatially non-uniform noise has been proposed. The noise subspace estimation problem is converted in the first phase of the proposed method to the problem of finding eigenvectors of a properly designed matrix so that the noise covariance matrix estimation is evoided. In the second phase, the covariance matrix of the non-uniform noise is first estimate based on the results of the first phase and then it is used for finding the noise subspace more accurately by means of generalized ED of this matrix and the data covariance matrix. It is of importance that the proposed method has low computational complexity and is non-iterative, and thus, has no issues with convergence. Moreover, it is superior to the existing iterative state-of-art methods in both the performance and especially the computational cost.

\ifCLASSOPTIONcaptionsoff
  \newpage
\fi

\end{document}